\documentclass[a4paper,12pt]{article}
\usepackage{amsmath}
\usepackage{amsfonts}
\usepackage[dvips]{graphicx}
\usepackage[bookmarksnumbered=true,breaklinks=true]{hyperref}
\usepackage{a4}
\usepackage{parskip}
\newcommand{\inv}[1]{\frac{1}{#1}}
\newcommand{\starco}[2]{\left[#1\stackrel{\star}{,}#2\right]}
\newcommand{\staraco}[2]{\left\{#1\stackrel{\star}{,}#2\right\}}
\newcommand{\co}[2]{\left[#1,#2\right]}
\newcommand{\var}[2]{\frac{\d #1}{\d #2}}
\newcommand{\vvar}[3]{\frac{\d^2 #1}{\d #2\d #3}}

\newcommand{\intx}{\int d^4x}

\newcommand{\Act}{S}
\newcommand{\ri}{{\rm i}}
\newcommand{\rig}{{\rm i}g}
\newcommand{\e}{\epsilon}
\newcommand{\s}{\sigma}
\renewcommand{\d}{\delta}
\renewcommand{\l}{\lambda}
\renewcommand{\th}{\theta}
\renewcommand{\k}{\tilde{k}}

\newcommand{\bc}{\bar{c}}
\newcommand{\bphi}{\bar{\phi}}

\newcommand{\hd}{\hat{\delta}}

\newcommand{\uim}{UV/IR mixing }
\newcommand{\nc}{non-commutative }
\title{\begin{flushright}{\small CERN-PH-TH/2007-085}\end{flushright}
\vspace{1cm}\bf{A Generalization of Slavnov-Extended Non-Commutative Gauge Theories}}
\author{\\[-0.3cm]\Large Daniel~N.~Blaschke\footnotemark[1]~ and Stefan~Hohenegger\footnotemark[2]~ }
\date{August 7, 2007}
\begin{document}
\maketitle
\begin{center}
\renewcommand{\thefootnote}{\fnsymbol{footnote}}
\vspace{-0.3cm}\footnotemark[1]Institute for Theoretical Physics, Vienna University of Technology\\Wiedner Hauptstrasse 8-10, A-1040 Vienna (Austria)\\[0.3cm]
\footnotemark[2]Department of Physics, CERN -- Theory Division\\CH-1211 Geneva 23 (Switzerland)\\[0.5cm]
\ttfamily{E-mail: blaschke@hep.itp.tuwien.ac.at, stefan.hohenegger@cern.ch}
\vspace{0.5cm}
\end{center}
\begin{abstract}
We consider a \nc U(1) gauge theory in 4 dimensions with a modified Slavnov term~\cite{Slavnov:2003} which looks similar to the 3-dimensional BF model. In choosing a space-like axial gauge fixing we find a new vector supersymmetry which is used to show that the model is free of \uim problems, just as in the previously discussed model~\cite{Blaschke:2006a,Blaschke:2006b}. Finally, we present generalizations of our proposed model to higher dimensions.
\end{abstract}
\newpage
\tableofcontents
\section{Introduction}
In general, \nc quantum field theories (NCQFT) realized through the Weyl-Moyal $\star$-product~\cite{Filk:1996}, suffer from UV/IR mixing, manifesting itself in the form of IR singularities for vanishing external momenta~\cite{Micu:2000}. Besides the possibility of performing a perturbative expansion in the deformation parameter of non-commutativity (see for example~\cite{Witten:1999,Bichl:2001,Bichl:2001d}), another way to get rid of this problem in the $U(1)$ gauge field sector was proposed by Slavnov~\cite{Slavnov:2003,Slavnov:2004}. It involves an extension of the gauge invariant action of the following form
\begin{align}\label{sl-term}
\intx\frac{\l}{2}\star\th^{\mu\nu}F_{\mu\nu},
\end{align}
introducing a new (dynamical) multiplier field $\l$.  The effect of this ''Slavnov term'' is such that the gauge field propagator of NCGFT becomes transversal with respect to $\k^\mu=\th^{\mu\nu}k_\nu$, where $\th^{\mu\nu}$ denotes the deformation parameter of \nc $3+1$ dimensional Minkowski space\footnote{In order to avoid problems with unitarity of the S-matrix and causality~\cite{Seiberg:2000,Connes:1997}, we choose $\th^{0i}=0$.}. Hence insertions of the (gauge independent) IR singular parts of the one-loop polarization tensor~\cite{Blaschke:2005b,Hayakawa:1999b,Ruiz:2000}
\begin{align}
\Pi^{\mu\nu}_{\text{IR}}(k)\sim \frac{\k^\mu\k^\nu}{(\k^2)^2},
\end{align}
are initially expected to vanish in higher order loop calculations. Unfortunately, though, new Feynman rules including the $\l$-field enter the model leading to new problems and divergent loop graphs. These new effects were discussed in~\cite{Blaschke:2005c} in great detail for a gauge fixing which interpolated between a covariant gauge and an axial gauge fixing.

For the case of a $\th_{\mu\nu}$ of reduced rank it was, however, shown in~\cite{Blaschke:2006a} upon choosing a special axial gauge (which allows for a gauge dependent linear vector supersymmetry (VSUSY) similar to the one of the 2-dimensional BF model~\cite{Schweda:1996,Schweda:1998,Gieres:2000,Blasi:2005b}) that the IR dangerous graphs do not yield any contribution, leading to the conclusion that the model is IR finite. Here, we wish to extend these results to more general $\th_{\mu\nu}$, by using a similar approach: we discuss a slight modification of the Slavnov term in order to incorporate properties of the 3-dimensional BF model~\cite{Schweda:1995,Schweda:1999,Piguet:1995} and again find a gauge in which we can conclude the absence of IR divergences.

The paper is organized as follows: In Section~\ref{vector-susy} we introduce the (gauge-fixed) action and find a great number of (gauge-dependent) symmetries, one of which is linear, fermionic and carries a vector index and which we will hence shortly call vector supersymmetry (VSUSY). This symmetry enables us in Section~\ref{loop-consequences} to essentially repeat the proof presented in reference~\cite{Blaschke:2006a} leading to the conclusion of IR-finiteness of the model. Finally, in Section~\ref{sec:gen-sl} we discuss the possibility of writing down topological-like terms in higher dimensions and comment on their consequences.

In order to simplify the notation, we will not spell out the star product symbol in the sequel: \emph{all products between fields (or functionals of fields) 
are understood to be star products}.

\section{The modified Slavnov term and symmetries of the action}\label{vector-susy}
\subsection{Action}\label{subsect:action}
The (gauge-)invariant action for a \nc $U(1)$ gauge field, enhanced by the extension proposed by Slavnov~\cite{Slavnov:2003}, is given by
\begin{align}\label{inv-act}
\Act_{\text{inv}}=&\intx\left[-\inv{4}F_{\mu\nu}F^{\mu\nu}+\frac{\l}{2}\th^{\mu\nu}F_{\mu\nu}\right],
\end{align}
where
\begin{align}
F_{\mu\nu}&=\partial_{\mu}A_{\nu}-\partial_{\nu}A_{\mu}-\rig
\left(A_{\mu}A_{\nu}-A_{\nu}A_{\mu}\right),\label{actinv}
\end{align}
denotes the field strength of the gauge connection and the signature of space-time is given by $g_{\mu\nu}=\text{diag}(+,-,-,-)$. In reference~\cite{Blaschke:2006a} the action (\ref{inv-act}) was interpreted as a topological 2-dimensional BF model coupled to Maxwell theory. The price, however, which had to be paid for this identification was a restriction of the (matrix-valued) parameter of non-commutativity to the special form,
\begin{align}
\th^{\mu\nu}\sim\left(\begin{array}{cccc} 0 & 0 & 0 & 0 \\ 0 & 0 & 1 & 0 \\ 0 & -1 & 0 & 0 \\ 0 & 0 & 0 & 0 \end{array}\right),
\end{align}
which made it possible to write the Slavnov term as $\frac{\l}{2}\e^{ab}F_{ab}$ with $a,b\in\{1,2\}$. In this section, however, we propose a possibility to consider a more general $\theta_{\mu\nu}$ without spoiling the topological nature of the theory. To this end we take $\th^{\mu\nu}$ to be completely arbitrary, at least in its spatial components\footnote{We assume the spatial coordinates commute with time in order to avoid various conceptional problems, as already mentioned.},
\begin{align}\label{theta-3dim}
\th^{\mu\nu}=\left(\begin{array}{cccc} 0 & 0 & 0 & 0 \\ 0 & 0 & \th^{12} & \th^{13} \\ 0 & -\th^{12} & 0 & \th^{23} \\ 0 & -\th^{13} & -\th^{23} & 0 \end{array}\right),
\end{align}
and remember that the Slavnov term was originally designed to introduce the following constraint on the field strength:
\begin{align}
\th^{12}F_{12}+\th^{13}F_{13}+\th^{23}F_{23}=0.\label{constraint}
\end{align}
We now impose the more restrictive constraint that each of the three terms vanishes by itself and implement this with the help of three multiplier fields $U_i(x)$ with $i\in\{1,2,3\}$ in the following way:
\begin{align}
\intx\left[-\inv{4}F_{\mu\nu} F^{\mu\nu}+U_3\th^{12}F_{12}+U_2\th^{13}F_{13}+U_1\th^{23}F_{23}\right].
\end{align}
Upon introducing the rescaled fields
\begin{align}
&\l_1\equiv \th^{23}U_1, &&\l_2\equiv -\th^{13}U_2, && \l_3\equiv \th^{12}U_3,
\end{align}
the invariant action can be rewritten in the form
\begin{align}
S_{\text{inv}}=&\intx\left[-\inv{4}F_{\mu\nu} F^{\mu\nu}+\inv{2}\e^{ijk}F_{ij}\l_k\right],\label{BFmod}
\end{align}
which is analogous to a 3-dimensional BF model coupled to Maxwell theory. Greek indices $\mu,\nu,\rho,\s$ take the values $0,1,2,3$ while Latin indices only denote the spatial directions $i,j,k,l\in\{1,2,3\}$. In fact, this action is invariant under two gauge symmetries. The first one is given by
\begin{align}\label{gauge1}
\d_{g1}A_\mu&=D_\mu\Lambda,\nonumber\\
\d_{g1}\l_k&=-\rig[\l_k,\Lambda],
\end{align}
and the second gauge symmetry reads
\begin{align}\label{gauge2}
\d_{g2}A_\mu&=0,\nonumber\\
\d_{g2}\l_k&=D_k\Lambda',
\end{align}
where $\Lambda$, $\Lambda'$ are gauge parameters. The covariant derivative $D_\mu$ is defined as
\begin{align}\label{defFD}
D_{\mu}\cdot=\partial_{\mu}\cdot-\rig\co{A_{\mu}}{\cdot}.
\end{align}
Observe, that $\Lambda'$ is a scalar and hence this model does not contain any so-called \emph{zero-modes}, which are typical for $n\geq4$-dimensional BF models (where $\Lambda'$ would be a $(n-3)$-form, cf.~\cite{Piguet:1993, Piguet:1995}). For the gauge fixing procedure we assume, that the algebra of fields is graded by the ghost-number and, accordingly, \emph{all commutators are considered to be graded} with respect to this degree, e.g. $\inv{2} {[c , c ]}$ stands for $\inv{2} \staraco{c}{c} = c \star c$ and ${[ A_{\mu} , c ]}$ stands for  $\starco{A_{\mu}}{c}$ $=$ $A_\mu\star c - c\star A_\mu$. 
At this point we would also like to draw attention to the fact that the deformation parameter $\th^{\mu\nu}$ does not appear explicitly in the Slavnov term of the action (\ref{BFmod}) (apart from its appearance in the star-products, of course). Therefore, it will make no difference which explicit form is chosen for $\th^{\mu\nu}$ in the upcoming considerations (i.e. we are free to chose any value for the entries $\th_{12}$, $\th_{13}$ and $\th_{23}$ in (\ref{theta-3dim})). The only restriction we need to take into account is that $\th^{0\mu}=0$ for reasons already mentioned.

We now continue by adding gauge fixing terms to our model in a BRST invariant way. To this end we 
fix both gauge symmetries using axial gauges following~\cite{Schweda:1995}:
\begin{align}\label{action1}
\Act=\intx\Big\{&-\inv{4}F_{\mu\nu} F^{\mu\nu}+\inv{2}\e^{ijk}F_{ij}\l_k+B n^iA_i+d m^i\l_i-\bc n^i D_ic\nonumber\\
&-\bphi m^i\big(D_i\phi-\rig\co{\l_i}{c}\big)\Big\}.
\end{align}
The multiplier fields $B$ and $d$ implement axial gauge fixings for the gauge symmetries (\ref{gauge1}) and (\ref{gauge2}), respectively. Both gauge fixings are chosen to be space-like ($n^0=m^0=0$) which we will find necessary in order to make the action invariant under a vector supersymmetry in the 3-dimensional subspace, as we will show in the next subsection. The remaining terms in (\ref{action1}) denote the ghost part of the action introducing the ghosts/antighosts $c,\bc,\phi,\bphi$. The canonical dimensions and ghost numbers for the various fields are summarized in Table~\ref{dimensions1}.
\begin{table}[ht]
\renewcommand{\arraystretch}{1.2}
\centering
\begin{tabular}{|c|c|c|c|c|c|c|c|c|}
\hline
& $A_\mu$ & $\l_k$ & $B$ &$ d$ & $c$ & $\bc$ & $\phi$ & $\bphi$\\\hline
dimension & 1 & 2 & 3 & 2 & 0 & 3 & 1 & 2 \\\hline
$\phi\pi$-charge & 0 & 0 & 0 & 0 & 1 & -1 & 1 & -1\\\hline
\end{tabular}
\caption{Canonical dimensions and ghost numbers of fields}
\label{dimensions1}
\renewcommand{\arraystretch}{1}
\end{table}

\noindent Before we discuss the symmetries of the action (\ref{action1}), let us consider the following: It is well-known in the literature (see e.g.~\cite{Schweda-book:1998} for a review), that axial gauge fixings render gauge theories ``ghost-free'', i.e. appropriate redefinitions of the multiplier fields implementing the gauge fixing lead to a decoupling of the ghost fields from the gauge fields. However, for us it will turn out to be convenient to merely decouple the ghosts from each other and choose $n^k=m^k$, as this will render the action invariant with respect to a \textbf{linear} vector supersymmetry. The necessary field redefinition is
\begin{align}\label{redef2}
d\quad\to\quad d'&=d-\rig\co{\bphi}{c}\,.
\end{align}
Hence, the action we will continue to work with is given by
\begin{align}\label{action}
\Act=\intx\Big\{&-\inv{4}F_{\mu\nu} F^{\mu\nu}+\inv{2}\e^{ijk}F_{ij}\l_k+B n^iA_i+d' m^i\l_i-\bc n^i D_ic-\bphi m^i D_i\phi\Big\},
\end{align}
with $n^k=m^k$.

\subsection{BRST \& VSUSY}\label{subsec:symm}
The action (\ref{action}) is invariant under the following BRST transformations, as can be easily verified:
\begin{align}\label{BRS}
& sA_\mu = D_\mu c, && s\l_i=D_i\phi-\rig\co{\l_i}{c}, \nonumber\\
& sc=\frac{\rig }{2}\co{c}{c}, && s\phi=\rig\co{\phi}{c}, \nonumber\\
& s \bar c = B, && s\bphi=d'+\rig\co{\bphi}{c}, \nonumber \\
& sB = 0, && sd'=-\rig\co{d'}{c},\nonumber \\
&s^2\varphi=0, \hspace{1cm} \text{for}&&\varphi\in\{A_{\mu},\l,B,d',c,\bc,\phi,\bphi\}.
\end{align}
The reason why $\bphi$ and $d'$ do not form a BRST-doublet similar to $\bc$ and $B$ lies in the field-redefinition $d'=d-\rig\co{\bphi}{c}$. However, this will be of no harm to us.

Furthermore, as already alluded to, the action is also invariant under the following fermionic symmetry
\begin{align}\label{true-susy}
& \d_iA_\mu=0, && \d_i\l_j=-\e_{ijk}n^k\bc, \nonumber\\
& \d_ic=A_i, && \d_i\phi=0, \nonumber\\
& \d_i\bc=0, && \d_i\bphi=0, \nonumber \\
& \d_iB=\partial_i\bc, && \d_id'=0,\nonumber \\
& \d_i\d_j\varphi=\d_0\varphi=0, \hspace{1cm} \text{for}&&\varphi\in\{A_{\mu},\l,B,d',c,\bc,\phi,\bphi\},
\end{align}
provided $n^k=m^k$. Besides the fact that the operator for this symmetry carries a space-time index, it is crucial to notice that it is a \textbf{linear} symmetry. In order to make contact with \cite{Blaschke:2006a} as well as with the (non-commutative) 3-dimensional BF model, we will hence refer to (\ref{true-susy}) as \textbf{vector supersymmetry} or VSUSY for short. The reason for the fact that the symmetry has a different form from the familiar one of BF models is obviously the presence of the $F_{\mu\nu}F^{\mu\nu}$-term in the action (\ref{action}) --- and of course the fact that we are dealing with $3+1$ dimensional space-time. As already anticipated, linearity of this symmetry was achieved through the field-redefinition $d'=d-\rig\co{\bphi}{c}$, while the initial multiplier field $d$ would have transformed non-linearly under VSUSY. However, linearity of the VSUSY will turn out to be crucial for our considerations.

The invariance of the action functional (\ref{action}) under the VSUSY-transformations (\ref{true-susy}) is described by the Ward identity
\begin{align}\label{Ward-id}
\mathcal{W}_i\Act\equiv\intx\left(\partial_i\bc\var{\Act}{B}+A_i\var{\Act}{c}+\e_{ijk}n^j\bc\var{\Act}{\l_k}\right)=0,
\end{align}
which will play an important role when considering loop corrections (cf. Section~\ref{loop-consequences}).

As we have seen, the VSUSY depends crucially on our choice of gauge. Moreover, the interplay of the form of $\th^{\mu\nu}$ (as given by equation (\ref{theta-3dim})) together with the space-like nature of the chosen gauge vector gives rise to even more symmetries as we are about to show right now. Let us take a look at the algebra satisfied by the BRST symmetry and the VSUSY: From relations (\ref{BRS}) and (\ref{true-susy}) it follows that\footnote{The equations of motion are displayed in Appendix~\ref{app:eom}.}
{\allowdisplaybreaks
\begin{subequations}\label{algebra}
\begin{align}
&\co{s}{s}\varphi=\co{\d_i}{\d_j}\varphi=0, \hspace{1cm} \text{for}\ \varphi=\{A_{\mu},\l_j,B,d',c,\bc,\phi,\bphi\},\label{algebra0}\\
&\co{s}{\d_i}A_j=\partial_iA_j-\e_{ijk}\var{\Act}{\l_k}+\hd_iA_j,\label{algebra1a}\\
&\co{s}{\d_i}A_0=\partial_iA_0+\hd_iA_0,\label{algebra1b}\\
&\co{s}{\d_i}c=\partial_ic,\label{algebra2}\\
&\co{s}{\d_i}\bc=\partial_i\bc,\label{algebra3}\\
&\co{s}{\d_i}B=\partial_iB,\label{algebra4}\\
&\co{s}{\d_i}\l_j=\partial_i\l_j-\e_{ijk}\var{\Act}{A_k}-D_i\var{\Act}{\l^j}+\hd_i\l_j,\label{algebra5}\\
&\co{s}{\d_i}d'=\partial_id'+\hd_id',\label{algebra6}\\
&\co{s}{\d_i}\phi=\partial_i\phi+\hd_i\phi,\label{algebra7}\\
&\co{s}{\d_i}\bphi=\partial_i\bphi+\hd_i\bphi,
\end{align}
\end{subequations}
implying a new bosonic vectorial symmetry of the action (\ref{action}) whose action on the fields is given by the transformation laws}
\begin{align}\label{3rdsymm}
\hd_iA_j&=\e_{ijk}n^kd'\,,\nonumber\\
\hd_iA_0&=-F_{i0}\,,\nonumber\\
\hd_i\l_j&=\e_{ijk}D_0F^{0k}-D_j\l_i+\inv{2}\e_{lmi}D_jF^{lm}+n_jD_id'-\rig\e_{ijk}n^k\co{\bphi}{\phi}\,,\nonumber\\
\hd_id'&=-D_id'\,,\nonumber\\
\hd_i\phi&=-D_i\phi\,,\nonumber\\
\hd_i\bar\phi&=-D_i\bphi\,,\nonumber\\
\hd_0\varphi&=0, \hspace{1cm} \text{for all fields }\varphi.
\end{align}
From the right hand side of (\ref{algebra}) we already see that the algebra of symmetries can only close \emph{on-shell}. Apart from the new symmetry (\ref{3rdsymm}) we also notice that the space translations $\partial_i$ appear. 

\subsection{Differences compared to the 2 dimensional BF-type Slavnov term}
In reference~\cite{Blaschke:2006a} it was shown that the algebra of BRST, VSUSY, the vectorial bosonic symmetry and translation symmetry closes on-shell for \nc Maxwell theory with a Slavnov term resembling the 2 dimensional BF model. Here, however, things are slightly more complicated:

Computing further commutators, we readily find that
\begin{align}
\co{s}{\hd_i}\varphi&=0,
\end{align}
for all fields. However, in trying to work out the complete symmetry algebra, one encounters more symmetries, e.g.
\begin{align}
\co{\d_i}{\hd_j}\l_k&=\e_{ijl}n^lD_k\bc\,,\nonumber\\
\co{\d_i}{\hd_j}c&=\hd_iA_j\,,\nonumber\\
\co{\d_i}{\hd_j}\varphi&=0, \hspace{1cm} \text{for all other fields }\varphi\,.
\end{align}
The right hand sides of these expressions represent new symmetry transformations of the action (\ref{action}), as can be easily checked. Similarly one obtains
\begin{align}
\co{\hd_i}{\hd_j}\varphi&=\text{further symmetry transf. of }\varphi\,.
\end{align}
In fact, computation of even more commutators between the new symmetries reveals numerous further ones which, however, shall not be discussed here. We are primarily interested in the linear vector supersymmetry denoted by $\d_i$ and will discuss its consequences in the next section.

But first we would like to draw attention to an interesting feature of the new bosonic vectorial symmetry (\ref{3rdsymm}): Inspired by its pendant in~\cite{Blaschke:2006a}, which was a symmetry of the gauge invariant action, we easily find the corresponding symmetry for the gauge invariant action (\ref{BFmod}) in our present model:
\begin{align}\label{symm1a}
&\hd_i^{(1a)}A_j=0,\nonumber\\
&\hd_i^{(1a)}A_0=-F_{i0}\nonumber\\
&\hd_i^{(1a)}\l_j=\e_{ijk}D_0F^{0k}.
\end{align}
In contrast to the situation in~\cite{Blaschke:2006a}, the gauge fixing of (\ref{action}) \emph{breaks} this symmetry. Instead (due to our space-like axial gauge fixing) it is replaced by\footnote{Notice, that the replacement (\ref{3rdsymm-1}) is not unique: The gauge fixed action (\ref{action}) is also invariant under $\hd'_i\l_j=\d_{ij}n^kD_kd'+\inv{2}\e_{lmi}D_jF^{lm}$ (where $\hd'_i\varphi=0$ for all other fields $\varphi$) and hence (\ref{3rdsymm-1}) might as well be replaced by an arbitrary linear combination of both, e.g. $\hd_i^{(1)}\to\hd_i^{(1)}-\hd'_i$.}
\begin{align}\label{3rdsymm-1}
&\hd_i^{(1)}A_j=\e_{ijk}n^kd',\nonumber\\*
&\hd_i^{(1)}A_0=-F_{i0}\nonumber\\*
&\hd_i^{(1)}\l_j=\e_{ijk}D_0F^{0k}-D_j\l_i+\inv{2}\e_{lmi}D_jF^{lm}+n_jD_id', \nonumber\\*
& \hd_i^{(1)}d'=-D_id', \nonumber\\*
&\hd_i^{(1)}\varphi=0, \hspace{1cm} \text{for}\hspace*{1cm}\varphi\in\{B,c,\bc,\phi,\bphi\}.
\end{align}
It is further amusing to see that the transformations $\hd_i^{(1b)}\l_j=-D_j\l_i$ and $\hd_i^{(1c)}\l_j=\inv{2}\e_{lmi}D_jF^{lm}$ leave the gauge invariant action (\ref{BFmod}) invariant as well. In fact, looking at $\hd_i^{(1b)}\l_j$ one is strongly reminded of the second gauge symmetry (\ref{gauge2}). The remaining field transformations of (\ref{3rdsymm}) form another symmetry\footnote{Remember, that the BRST transformations were already made up of two separate symmetries, namely those corresponding to the two gauge symmetries (\ref{gauge1}) and (\ref{gauge2}).} of the gauge fixed action (\ref{action}) which does not involve the gauge field $A_\mu$:
\begin{align}\label{3rdsymm-2}
&\hd_i^{(2)}\l_j=-\rig\e_{ijk}n^k\co{\bphi}{\phi},\nonumber\\*
&\hd_i^{(2)}\phi=-D_i\phi,\nonumber\\*
&\hd_i^{(2)}\bphi=-D_i\bphi,\nonumber\\*
&\hd_i^{(2)}\varphi=0, \hspace{1cm} \text{for}\hspace*{1cm}\varphi\in\{A_\mu,c,\bc,B,d'\}\,.
\end{align}
So in contrast to the simpler model of \nc Maxwell theory with a Slavnov term resembling the 2 dimensional BF model, the right hand sides of the commutators $\co{s}{\d_i}$ reveal a linear combination of \emph{two} symmetries ($\hd_i=\hd_i^{(1)}+\hd_i^{(2)}$), one of which is a modified version of (\ref{symm1a}) due to gauge fixing, namely (\ref{3rdsymm-1}). Furthermore, the algebra does not close immediately, but instead, many additional symmetries appear.

In conclusion of this subsection: The appearance of an additional bosonic vectorial symmetry of the gauge invariant action seems to be typical for Yang Mills theories with a BF-type Slavnov term. However, its survival after gauge fixing is in general not compatible with the existence of a linear VSUSY.

\section{Consequences of the vector supersymmetry}\label{loop-consequences}
The generating functional $Z^c$ of the connected Green functions is given by the Legendre transform of the generating functional $\Gamma$ of the one-particle irreducible Green functions
\begin{align}
Z^c=\Gamma+\int d^4x\left(j_A^\mu A_\mu+j_BB+j^i_\l\l_i+j_{d'}d'+ j_{c}c+j_{\bc}\bc+j_{\phi}\phi+j_{\bphi}\bphi\right),
\end{align}
where in the classical approximation $\Gamma$ essentially equals the action $S$. This leads to the usual relations
\begin{align}
&\var{Z^c}{j_A^\mu}=A_\mu,&&\var{Z^c}{j_B}=B,&&\var{Z^c}{j_\l^j}=\l_j,&&\var{Z^c}{j_{d'}}=d',\nonumber\\
&\var{Z^c}{j_c}=c,&&\var{Z^c}{j_{\bc}}=\bc,&&\var{Z^c}{j_\phi}=\phi,&&\var{Z^c}{j_{\bphi}}=\bphi,\nonumber\\
&\var{\Gamma}{A_\mu}=-j_A^\mu,&&\var{\Gamma}{B}=-j_B,&&\var{\Gamma}{\l_j}=-j_\l^j,&&\var{\Gamma}{d'}=-j_{d'},\nonumber\\
&\var{\Gamma}{c}=j_c,&&\var{\Gamma}{\bc}=j_{\bc},&&\var{\Gamma}{\phi}=j_\phi,&&\var{\Gamma}{\bphi}=j_{\bphi}.
\end{align}
In the tree graph approximation the Ward identity (\ref{Ward-id}) describing the linear vector supersymmetry in terms of $Z^c$ is given by
\begin{align}\label{ward2}
\mathcal{W}_iZ^c=\int d^4x\Big[& j_B\partial_i\var{Z^c}{j_{\bc}}-j_{c} \var{Z^c}{j_A^i} +\e_{ijk}n^j j_\l^k\var{Z^c}{j_{\bc}}\Big]=0.
\end{align}
Varying (\ref{ward2}) with respect to the appropriate sources one gets the following relations:
\begin{subequations}\label{rel-all}
\begin{align}
\vvar{Z^c}{j_A^i}{j_\l^j}\Bigg|_{j=0}&=\e_{ijk}n^k\vvar{Z^c}{j_{\bar{c}}}{j_c}\Bigg|_{j=0}\,,\label{rel-ghost-mixed}\\
\vvar{Z^c}{j_A^i}{j_A^\nu}\Bigg|_{j=0}&=0\,.\label{rel-Aij}
\end{align}
\end{subequations}
Furthermore, one has the gauge fixing conditions (cf. (\ref{action}), $m^i=n^i$)
\begin{subequations}
\begin{align}
-j_B&=n^i\var{Z^c}{j^i_A},\\
-j_{d'}&=n^i\var{Z^c}{j^i_\l},
\end{align}
and the (anti)ghost equations
\begin{align}
-n^i\partial_i\var{Z^c}{j_{\bc}}-\rig\co{j_B}{\var{Z^c}{j_{\bc}}}&=j_{c},\quad\qquad-n^i\partial_i\var{Z^c}{j_{c}}-\rig\co{j_B}{\var{Z^c}{j_{c}}}&=j_{\bc}\,,\label{ghost-eq1}\\
-n^i\partial_i\var{Z^c}{j_{\bphi}}-\rig\co{j_B}{\var{Z^c}{j_{\bphi}}}&=j_{\phi},\quad\qquad-n^i\partial_i\var{Z^c}{j_{\phi}}-\rig\co{j_B}{\var{Z^c}{j_{\phi}}}&=j_{\bphi}\,,\label{ghost-eq2}
\end{align}
\end{subequations}
from which follow
\begin{subequations}\label{rel-other}
\begin{align}
n^i\vvar{Z^c}{j_B(y)}{j_A^i(x)}\Bigg|_{j=0}&=-\d^4(x-y),\label{rel-AB-prop}\\
n^i\vvar{Z^c}{j_{d'}(y)}{j_{\l}^i(x)}\Bigg|_{j=0}&=-\d^4(x-y),\label{rel-dl-prop}\\
n^i\partial_i\vvar{Z^c}{j_c(y)}{j_{\bc}(x)}\Bigg|_{j=0}&=-\d^4(x-y),\label{rel-ghost-eq1}\\
n^i\partial_i\vvar{Z^c}{j_{\phi}(y)}{j_{\bphi}(x)}\Bigg|_{j=0}&=-\d^4(x-y).\label{rel-ghost-eq2}
\end{align}
\end{subequations}
In momentum space, the free propagators of the theory with $m^i=n^i$ are given by (see Appendix~\ref{app-propagators})
\begin{subequations}\label{propagators}
\begin{align}
&\ri\Delta^{c\bc}(k)=-\inv{(nk)},\qquad \ri\Delta^{\phi\bphi}(k)=-\inv{(nk)},\\ &\ri\Delta^{AB}_i(k)=\frac{\ri k_i}{(nk)},\qquad \ri\Delta^{d'\l}_i(k)=\frac{\ri k_i}{(nk)},\\
&\ri\Delta_{ij}^{\l\l}(k)=\frac{-k^2}{\vec{k}^2}\left(g_{ij}-\frac{k_in_j+n_ik_j}{(nk)}+n^2\frac{k_ik_j}{(nk)^2}\right),\label{propagatorsl}\\
&\ri\Delta_{ij}^{A\l}(k)=\frac{-\ri}{\vec{k}^2}\left(\e_{ilj}k^l-\e_{ilr}\frac{k^ln^rk_j}{(nk)}+\e_{jlr}\frac{k^ln^rk_i}{(nk)}\right),\\
&\ri\Delta_{i0}^{A\l}(k)=\frac{-\ri}{\vec{k}^2}\left(-\e_{ilr}\frac{k^ln^rk_0}{(nk)}\right),\\
&\ri\Delta_{00}^{AA}(k)=-\inv{k^2}\left(g_{00}-\frac{k_0^2}{\vec{k}^2}\right)=\inv{\vec{k}^2},\quad \ri\Delta_{ij}^{AA}(k)=\ri\Delta_{i0}^{AA}(k)=0,\label{propagatorsA}
\end{align}
\end{subequations}
and one easily sees that the relations (\ref{rel-all}) and (\ref{rel-other}) hold\footnote{$\k^2=-(k_1^2+k_2^2)$, $(nk)=-(n_1k_1+n_2k_2)$, $(n\k)=(n_1k_2-n_2k_1)$, and similarly for $n^i\leftrightarrow m^i$. Furthermore, $i\Delta^{\varphi_1\varphi_2}(x-y)=-i\vvar{Z^c}{j_{\varphi_1}(x)}{j_{\varphi_2}(y)}\Big|_{j=0}$ for all fields $\varphi$.}. Furthermore, by virtue of equation (\ref{propagatorsA}) and $\th^{\mu0}=0$, the gauge field propagator is still transverse with respect to $\k_\mu\equiv\th_{\mu\nu}k^\nu$ despite the modification of the Slavnov term (cf. (\ref{sl-term}) and (\ref{action})).

Finally, the vector supersymmetry leads to the following nice features for loop calculations: Obviously, the combination of the $\l A$-vertex $V_{ijk}^{\l A}\propto \e_{ijk}$ with a gauge field propagator $\Delta^{AA}_{\mu\nu}$ is always zero (see eq. (\ref{rel-Aij})). But since it is impossible to have $\l A$-vertices in arbitrary loop graphs (except for tree graphs) unless some of them couple to gauge field propagators~\cite{Blaschke:2006a}, such graphs will not contribute to any quantum corrections. \emph{Hence, neither the $\l$-vertex nor the $\l$/$\l A$-propagators contribute to the gauge field self-energy corrections at any loop-order!} In union with the transversality condition of the gauge field propagator, it therefore follows that no IR divergences from 1-loop graph insertions are passed on to higher loop orders.

Note, that we have only discussed the IR behaviour of our model and the UV sector, especially the planar graphs, remain to be thoroughly analyzed. Due to the VSUSY the $\l$-field does not play a role in the UV sector either and therefore we do not expect any major problems. Still one needs to take care when computing the Feynman graphs due to the axial gauge fixing, i.e. an appropriate prescription for the $(nk)^{-1}$ poles is needed (see for example~\cite{Leibbrandt:1994} and references therein).

\section{Generalization to arbitrary dimensions}\label{sec:gen-sl}
\subsection{Re-interpretation of the action}
In Section \ref{subsect:action}, we modified the original Slavnov term proposed in \cite{Slavnov:2003,Slavnov:2004} by changing the scalar field $\l$ into a set of fields $\l_i$, labelled by an index corresponding to the \nc subsector of space-time. In order to show that the Slavnov trick works we have taken a rather pragmatic point of view and have not inquired further about the true nature of $\l_i$. In fact an intriguing observation can be made when returning to the action (\ref{action}) and explicitly writing out the field strength $F_{\mu\nu}$ in the Slavnov term:
\begin{align}\label{ChernSimons2}
\Act=\intx\Big\{&-\inv{4}F_{\mu\nu} F^{\mu\nu}+\e^{ijk}\l_i\partial_jA_k-\rig\e^{ijk}\l_i A_j A_k+B n^iA_i+d' m^i\l_i-\nonumber\\
&-\bc n^i D_ic-\bphi m^i D_i\phi\Big\}.
\end{align}
Written in this way, the generalized Slavnov term has certain similarities with a Chern-Simons type term if $\l_i$ is interpreted as a second gauge field. In order to make this observation even more striking, we rescale the fields according to
\begin{align}
&\l_i\equiv \mu\l'_i,&&d'\equiv \frac{d^{\prime\prime}}{\mu},
\end{align}
where $\mu$ is a constant with mass dimension 1. For the action, we then find
\begin{align}
\Act=\intx\Big\{&-\inv{4}F_{\mu\nu} F^{\mu\nu}+\mu\e^{ijk}\l'_i\partial_jA_k-\rig\mu\e^{ijk}\l'_i A_j A_k+B n^iA_i+d^{\prime\prime} m^i\l'_i-\nonumber\\
&-\bc n^i D_ic-\bphi' m^i D_i\phi'\Big\}.\label{intrgauge}
\end{align}
Note, that $\phi'$ and $\bphi'$ differ from $\phi$ and $\bphi$ by their canonical dimension, which can be seen from Table~\ref{dimensionsnew}.
\begin{table}[ht]
\renewcommand{\arraystretch}{1.2}
\centering
\begin{tabular}{|c|c|c|c|c|c|c|c|c|}
\hline
& $A_\mu$ & $\l'_k$ & $B$ &$ d^{\prime\prime}$ & $c$ & $\bc$ & $\phi'$ & $\bphi'$\\\hline
dimension & 1 & 1 & 3 & 3 & 0 & 3 & 0 & 3 \\\hline
$\phi\pi$-charge & 0 & 0 & 0 & 0 & 1 & -1 & 1 & -1\\\hline
\end{tabular}
\caption{Canonical dimensions and ghost numbers of redefined fields}
\label{dimensionsnew}
\renewcommand{\arraystretch}{1}
\end{table}

\noindent
Thus the two sets of fields $(A_\mu,B,c,\bc)$ and $(\l'_i,d^{\prime\prime},\phi',\bphi')$ not only appear in a rather similar way in the action, but also their canonical dimensions match precisely. This provides some evidence that $(\l'_i,d^{\prime\prime},\phi',\bphi')$ should really be interpreted as another gauge field together with a second ghost system. In addition to the classical dynamics a striking difference is the absence of a $\l'_0$ component. Indeed, $\l'_i$ has only components corresponding to 
potentially 
\nc directions. As we will see, this is a general feature when considering similar examples in a different number of dimensions, as we will do in the next section.

\subsection{Topological terms in higher dimensions}
In reference~\cite{Blaschke:2006a} it was shown that the interpretation of the Slavnov-term as a topological-type term (resembling a 2 dimensional BF model) is fruitful in studying the fate of the IR divergences in more detail. Also in Section \ref{loop-consequences}, we have encountered that modifying the Slavnov term to resemble a 3 dimensional BF model teaches us interesting lessons in this respect. In doing so, however, we had to add an index to the $\l$ field, which (as we have just seen) allows for interpreting it in terms of a gauge field. It is expected that increasing the dimension of the non-commutative subspace (which necessarily also involves increasing the dimension of space-time) will lead to objects with yet more indices whose interpretations remain to be seen. Therefore, besides being interesting in its own right, we might learn a good deal about $\l$ (whatever its ``form degree'' might be), by introducing Slavnov terms in higher dimensions, which can again be interpreted as being topological in the same sense as before.

To this end, consider a $D>2$ dimensional space-time $\mathcal{M}$, which we write as the product of a $(D-n)$-dimensional Minkowski space and a $n$-dimensional \nc Euclidean space
\begin{align}\label{D-n-dim}
\mathcal{M}=\mathbb{M}_{D-n}\times \mathbb{R}_n^{\text{NC}}.
\end{align}
We restrict $n$ to be $2\leq n<D$, since we want to have at least two \nc dimensions and we furthermore want to interpret one dimension as time. In accordance with Section \ref{vector-susy}, space-time indices of the whole $\mathcal{M}$ are denoted by Greek letters, $\mu,\nu\in\{0,1,\ldots, D-1\}$, while the \nc directions are labelled by Latin indices $i,j\in\{D-n,\ldots ,D-1\}$. In this setup, the analog to the constraint (\ref{constraint}) is a sum of $\frac{n(n-1)}{2}$ terms and in the following we will impose the stronger demand that each of them vanishes separately. Let us consider this in somewhat more detail:

\noindent{\underline{$\mathbf{D=3:}$}}\\
In this simplest case, the only possibility is to choose $n=2$, which renders $\th^{\mu\nu}$ of the form
\begin{align}
\th^{\mu\nu}=\left(\begin{array}{ccc} 0 & 0 & 0 \\ 0 & 0 & \th \\ 0 & -\th & 0 \end{array}\right),\hspace{1cm}\text{with}\ \th\neq 0.
\end{align}
The transversality constraint (\ref{constraint}) consists of a single term
\begin{align}
\th F_{12}=0,
\end{align}
which is implemented in the action by a scalar field $\l$:
\begin{align}
\int d^3x\l\th^{\mu\nu}F_{\mu\nu}=\int d^3x\l\th\e^{ij}F_{ij}.
\end{align}

\noindent{\underline{$\mathbf{D=4:}$}}\\
Here there are two possibilities for $n$, namely 2 and 3, as can be seen in the following table
\begin{center}
\begin{tabular}{c|c|c|c|c}
$n$ & $\th^{\mu\nu}$ & \textbf{constraints} & $\l$-\textbf{field} & \textbf{action term}\\\hline 
$2$ & $\left(\begin{array}{cccc} 0 & 0 & 0 & 0 \\ 0 & 0 & 0 & 0 \\ 0 & 0 & 0 & \th \\ 0 & 0 & -\th & 0 \end{array}\right)$ & $F_{23}=0$ & $\l$ & \parbox{3.5cm}{\begin{align}\int d^4x\l\th^{\mu\nu}F_{\mu\nu}\propto\nonumber\\
\propto\int d^4x\l\e^{ij}F_{ij}\nonumber\end{align}} \\\hline
$3$ & $\left(\begin{array}{cccc} 0 & 0 & 0 & 0 \\ 0 & 0 & \th^{12} & \th^{13} \\ 0 & -\th^{12} & 0 & \th^{23} \\ 0 & -\th^{13} & -\th^{23} & 0 \end{array}\right)$ & \parbox{2.5cm}{\begin{align}&\th^{12} F_{12}=0\nonumber \\ & \th^{13}F_{13}=0 \nonumber\\ &\th^{23}F_{23}=0\nonumber\end{align}} & $\l_{i}$ & \parbox{3.5cm}{\begin{align}\int d^4x\e^{ijk}F_{ij}\l_k\nonumber\end{align}} \\
\end{tabular}
\end{center}
where $\th^{ij}\neq 0$ for all $i\neq j$. In the case $n=2$ (which is essentially the one studied in \cite{Blaschke:2006a}), $\l$ is obviously a scalar once more, while for $n=3$, $\l_i$ enjoys the interpretation as a vector field with components only in the $\mathbb{R}_3^{\text{NC}}$, as we have already pointed 
out\footnote{In general, the number of Lagrange multipliers $\l_i$ might as well be greater than the number of non-vanishing $\th^{ij}$. However, in this section we are primarily interested in the case where they are equal.}.

\noindent\underline{\textbf{Generic $\mathbf{D}$:}}\\
From the two previous examples, we can easily generalize the case of generic $D$ and $n$: Let us again start out with the most generic $\th^{\mu\nu}$ 
\begin{align}
\th^{\mu\nu}=\left(\begin{array}{ccc} 0 & \\ & \th^{ij}\end{array}\right),\label{fullmatt}
\end{align}
with $\th^{ij}\neq 0 $ for all $i\neq j$. The Slavnov constraint one has to impose on this model reads
\begin{align}
\th^{ij}F_{ij}=0,\hspace{1cm}\text{with }\, D-n\leq i<j\leq D-1,\label{heavyconstraint}
\end{align}
and the stronger constraints, where each term in the sum is zero, are implemented with the help of $\frac{n(n-1)}{2}$ Lagrange multipliers, which can be arranged into a field $\l_{i_1\ldots i_{n-2}}$ which is totally antisymmetric in all its indices. The corresponding action term is of the form
\begin{align}
\int d^Dx\e^{ijk_1\ldots k_{n-2}}F_{ij}\l_{k_1\ldots k_{n-2}},\label{acttermfull}
\end{align}
resembling a $n$ dimensional BF model (see e.g.~\cite{Piguet:1995,Piguet:1993,Maggiore:1992}). Note, that the field $\l_{i_1\ldots i_{n-2}}$ has a convenient interpretation as a $(n-2)$ form which only has components in $\mathbb{R}_n^{\text{NC}}$.

We would, however, like to stress the following points:
\begin{itemize}
\item although we started with the parameter matrix of non-commutativity (\ref{fullmatt}) with $\th^{ij}\neq 0 $ for all $i\neq j$ to give the Slavnov constraint a suggestive form, the action term (\ref{acttermfull}) is in principle valid for any choice of the $\theta_{ij}$,
\item we choose the maximum number of constraints compatible with the Slavnov trick.
\end{itemize}

\noindent
Before closing this subsection, let us comment on a special case where we set some of the $\theta_{ij}=0$ in (\ref{fullmatt}) in a rather peculiar way and see if we find alternatives to the constraints (\ref{acttermfull}) resembling topological terms. We hence consider the matrix $\th^{\mu\nu}$ having the block-diagonal structure
\begin{align}
\th^{\mu\nu}=\left(\begin{array}{cccc}0_{D-n} & & & \\ & \th_{n_1} & & \\ & & \ddots & \\ & & & \th_{n_p}\end{array}\right), \hspace{1cm}\text{with}\ \sum_{a=1}^pn_a=n,
\end{align}
where $0_{D-n}$ stands for a $(D-n)\times (D-n)$ square matrix with 0 entries everywhere, and $\th_{n_a}$ are antisymmetric $n_a\times n_a$ matrices (with $2\leq n_a\leq n$) with non-zero off-diagonal entries. In other words, we consider a space with $p$ \nc subspaces, which, however, commute among each other.

If we now label the indices of the $a$-th \nc block\footnote{They take values $D-n-1+\sum_{b=1}^{a-1}n_b< i^{(a)}< D-n+\sum_{b=1}^{a}n_b$.} by $i^{(a)}$, we can impose the following set of (alternative) constraints
\begin{align}
&\th^{i^{(1)}_1i^{(1)}_2}F_{i^{(1)}_1i^{(1)}_2}=0,\quad\text{with }\, D-n-1< i^{(1)}_1<i^{(1)}_2< D-n+n_1\,,\nonumber\\
&\hspace{1.5cm}\vdots \nonumber\\
&\th^{i^{(a)}_1i^{(a)}_2}F_{i^{(a)}_1i^{(a)}_2}=0,\quad\text{with }\, D-n-1+\sum_{b=1}^{a-1}n_b< i^{(a)}_1<i^{(a)}_2< D-n+\sum_{b=1}^{a}n_b\,,\label{indexrange}\nonumber\\
&\hspace{1.5cm}\vdots \nonumber\\
&\th^{i^{(p)}_1i^{(p)}_2}F_{i^{(p)}_1i^{(p)}_2}=0,\quad\text{with }\, D-n-1+\sum_{b=1}^{p-1}n_b< i^{(p)}_1<i^{(p)}_2< D-n+\sum_{b=1}^{p}n_b\,,
\end{align}
where we suspended summation over repeated indices. These constraints suggest to consider the following term in the action
\begin{equation}\label{gen-sl}
\sum_{a=1}^p\int d^Dx\e^{i^{(a)}_1\ldots i^{(a)}_{n_a}}\l^{(a)}_{i^{(a)}_1\ldots i^{(a)}_{n_a-2}}F_{i^{(a)}_{n_a-1},i^{(a)}_{n_a}}\,,
\end{equation}
which can be interpreted as a sum of $n_a$ dimensional BF terms and the $\l^{(a)}_{i^{(a)}_1\ldots i^{(a)}_{n_a-2}}$ can be identified as $(n_a-2)$ forms with components in the $a$-th \nc subspace. The symbol $\e^{i^{(a)}_1\ldots i^{(a)}_{n_a}}$ is defined similarly to the Levi-Civita symbol, i.e. it is $+1$ ($-1$) for even (odd) permutations of its indices. The only difference here is that the range of the indices $i^{(a)}_l$ is given by (\ref{indexrange}) rather than being $1,\ldots,n_a$.

It is also important to stress that the superscript ``$(a)$'' of the $\l^{(a)}_{i^{(a)}_1\ldots i^{(a)}_{n_a-2}}$  is not an index but only a label for the various multiplier fields.

\subsection{Generalized Slavnov terms and VSUSY}\label{subsec:sl-vsusy}
After having gained some intuitive understanding of the nature of the $\l$ field and having generalized the actions considered in \cite{Blaschke:2006a} as well as in Section \ref{vector-susy}, we might now ask which further notions we are able to generalize to higher dimensions. One interesting point is what happens to the VSUSY in higher dimensions.

We have seen that the action (\ref{action}) is invariant under the vector supersymmetry described by (\ref{Ward-id}). On the other hand, if one replaces the gauge invariant part of (\ref{action}) with (\ref{inv-act}) and (\ref{theta-3dim}), hence implementing the weaker Slavnov constraint (\ref{constraint}), one cannot find VSUSY. A first step is therefore to make clear, if we can find a gauge fixing, so that an action including Slavnov terms of the form (\ref{gen-sl}) becomes invariant under a vector supersymmetry. From all we know so far, such a gauge fixing has to be of an axial type. Let us consider a simple example, namely $(D=5,n=4)$ and a parameter of non-commutativity of the form
\begin{align}
\th^{\mu\nu}=\left(\begin{array}{ccccc} 0 & 0 & 0 & 0 & 0 \\ 0 & 0 & \th_1 & 0 & 0 \\  0 & -\th_1 & 0 & 0 & 0 \\  0 & 0 & 0 & 0  & \th_2 \\ 0 & 0 & 0 & -\th_2 & 0\end{array} \right),
\end{align}
and the following (gauge fixed) action
\begin{align}
\Act=\int d^5x\bigg(&-\inv{4}F_{\mu\nu}F^{\mu\nu}+\frac{\l^{(1)}}{2}\e^{i^{(1)}j^{(1)}}F_{i^{(1)}j^{(1)}}+\frac{\l^{(2)}}{2}\e^{i^{(2)}j^{(2)}}F_{i^{(2)}j^{(2)}}+Bn^\mu A_\mu-\nonumber\\
&-\bc n^\mu D_\mu c\bigg).
\end{align}
We choose $n^\mu$ to only have spatial components ($n^0=0$). The action is BRST invariant with
\[
s\l^{(1,2)}=-\rig\co{\l^{(1,2)}}{c},
\]
and the transformations of the other fields given by (\ref{BRS}). Since the two Slavnov terms represent 2-dimensional BF terms in the $x_1,x_2$-plane and the $x_3,x_4$-plane respectively, one could na\"{\i}vely assume invariance of the action under the following VSUSY transformations:
\begin{align}
&\d_\mu A_\nu=\d_\mu\bc=0, &&\d_{i^{(1)}}\l^{(1)}=\e_{i^{(1)}j^{(1)}}n^{j^{(1)}}\bc,\nonumber\\
&\d_{i^{(1)}}c=A_{i^{(1)}}, && \d_{i^{(2)}}\l^{(1)}=0,\nonumber\\
&\d_{i^{(2)}}c=A_{i^{(2)}}, && \d_{i^{(1)}}\l^{(2)}=0,\nonumber\\
&\d_{i^{(1)}}B=\partial_{i^{(1)}}\bc, && \d_{i^{(2)}}\l^{(2)}=\e_{i^{(2)}j^{(2)}}n^{j^{(2)}}\bc,\nonumber\\
&\d_{i^{(2)}}B=\partial_{i^{(2)}}\bc, && \d_0\varphi=0\quad\text{for all fields.}
\end{align}
However, direct calculations show that
\begin{align}
\d_{i^{(1)}}\Act&=\int d^5x\left(\bc n^{j^{(2)}}F_{j^{(2)}i^{(1)}}\right)\neq0,\nonumber\\
\d_{i^{(2)}}\Act&=\int d^5x\left(\bc n^{j^{(1)}}F_{j^{(1)}i^{(2)}}\right)\neq0.
\end{align}
So obviously, we have invariance under $\d_{i^{(1)}}$ if we choose $n^{j^{(2)}}=0$ or invariance under $\d_{i^{(2)}}$ if we choose $n^{j^{(1)}}=0$ but never under both. For higher dimensional models with arbitrary Slavnov terms of the type (\ref{gen-sl}) it therefore makes sense to assume that, depending on the choice of the axial gauge fixing vector $n^\mu$, one can at most have invariance under a vector supersymmetry whose operator acts non-trivially \emph{only in one} of the $n_a$-dimensional subspaces corresponding to the $a$-th BF term.

In fact, the transformations for VSUSY in the $a$-th non-commutative subspace (i.e. the $a$-th summand in equation (\ref{gen-sl})) of an arbitrary dimensional BF-Slavnov model are always the same, namely the only non-trivial transformations are\footnote{This of course includes the case $p=1$ in (\ref{gen-sl}).}
\begin{align}\label{na-VSUSY}
&\d_{i^{(a)}}c=A_{i^{(a)}},&&\d_{i^{(a)}}\l^{(a)}_{j^{(a)}_1\cdots j^{(a)}_{n_a-2}}=\e_{i^{(a)}k^{(a)}j^{(a)}_1\cdots j^{(a)}_{n_a-2}}n^{k^{(a)}}\bc,\nonumber\\
&\d_{i^{(a)}}B=\partial_{i^{(a)}}\bc,
\end{align}
with the range of indices given in (\ref{indexrange}). For the sake of clarity we will drop the superscripts ``$(a)$'' in the following and keep in mind that we are referring to the $a$-th BF term. The linear VSUSY (\ref{na-VSUSY}) exists only after appropriate redefinition of the multiplier fields fixing the gauge symmetries: Let the collection of $2(n_a-2)$ fields
\begin{align}
\left\{\phi,\phi_{j_1},\phi_{j_1j_2},\ldots,\phi_{j_1\ldots j_{n_a-3}}\right\}\,,\nonumber\\
\left\{\bar{\phi},\bar{\phi}_{j_1},\bar{\phi}_{j_1j_2},\ldots,\bar{\phi}_{j_1\ldots j_{n_a-3}}\right\}\,,
\end{align}
be the tower of ghosts\footnote{See for example~\cite{Piguet:1995,Piguet:1993,Maggiore:1992} and references therein.} we need to introduce. For $n_a=2$ no ghosts are needed since $\l$ is a scalar in that case. Furthermore, let the $n_a-2$ objects 
\begin{align}
\left\{d,d_{j_1},\ldots ,d_{j_1\ldots j_{n_a-3}}\right\}
\end{align}
be Lagrange multipliers fixing the gauge freedom of\footnote{There is no gauge freedom for the scalar $\phi$.} 
\begin{align}
\left\{\l_{j_1\ldots j_{n_a-2}},\phi_{j_1},\ldots,\phi_{j_1\ldots j_{n_a-3}}\right\}.
\end{align}
In order to have a linear VSUSY we must redefine the multipliers $d$ according to
\begin{align}
&d'=d-\rig\co{\bar{\phi}}{c},\nonumber\\
&d'_{j_1\ldots j_{m_a}}=d_{j_1\ldots j_{m_a}}-\rig\co{\bar{\phi}_{j_1\ldots j_{m_a}}}{c},\hspace{1cm} \forall\ 1\leq m_a\leq n_a-3.
\end{align}
leading to the BRST transformations\footnote{Concerning the BRST transformations for the other fields we refer to the literature~\cite{Piguet:1995,Piguet:1993,Maggiore:1992} once again.}
\begin{align}\label{na-BRS}
&s\bar{\phi}=d'+\rig\co{\bar{\phi}}{c},\nonumber\\
&s\bar{\phi}_{j_1\ldots j_{m_a}}=d'_{j_1\ldots j_{m_a}}+\rig\co{\bar{\phi}_{j_1\ldots j_{m_a}}}{c}, \hspace{1cm}\forall\ 1\leq m_a\leq n_a-3,\nonumber\\
&sd'=-\rig\co{d'}{c},\nonumber\\
&sd'_{j_1\ldots j_{m_a}}=-\rig\co{d'_{j_1\ldots j_{m_a}}}{c}, \hspace{1cm}\forall\ 1\leq m_a\leq n_a-3.
\end{align}
We should also stress that the vector supersymmetry operator (\ref{na-VSUSY}) acts non-trivially only on the $a$-th Slavnov term and the gauge fixing part for the gauge field $A_i$ of the action, provided, of course, its axial gauge fixing vector is chosen to be non-zero only in the $n_a$ dimensional subspace where it is identical to the axial gauge fixing vector for $\l^{(a)}_{j_1\cdots j_{n_a-2}}$.

Obviously, we would not completely loose VSUSY if we wrote the gauge fixing part of the action in terms of $d$ rather than $d'$, but the VSUSY would become \emph{non-linear}, e.g. the following non-linear VSUSY transformations would have to be added to (\ref{na-VSUSY}):
\begin{align}
&\d_{i}d=\rig\co{A_{i}}{\bar{\phi}},\nonumber\\
&\d_{i}d_{j_1\ldots j_{m_a}}=\rig\co{A_{i}}{\bar{\phi}_{j_1\ldots j_{m_a}}}, \hspace{1cm}\forall\ 1\leq m_a\leq n_a-3,
\end{align}
for all Lagrange multipliers.

An important point to mention, however, is that the presence of a linear vector supersymmetry alone is not sufficient to guarantee the complete absence of all IR divergences in the loop calculations. In fact, since we have found the VSUSY to act non-trivially only in a certain subspace of the \nc space, the argument at the end of Section \ref{loop-consequences} cannot be applied here which means we are not able to prove IR finiteness of the model in this way.

\section{Conclusions}
Inspired by recent results concerning Slavnov-extended gauge theories~\cite{Blaschke:2006a}, we discussed a step by step generalization of the Slavnov term. In Sections~\ref{vector-susy} and~\ref{loop-consequences} we considered the more restrictive version (\ref{BFmod}) of the Slavnov term resembling a 3 dimensional BF model. We found numerous new symmetries of the gauge fixed action, one of which is (\ref{true-susy}), a \emph{linear vector supersymmetry} (VSUSY) which (although it is gauge dependent and hence non-physical) allowed us to show that the model is free of quadratic IR divergences.

Section~\ref{sec:gen-sl} was then dedicated to possible generalizations to higher dimensional space-times of the form (\ref{D-n-dim}). We could show that in a specific setup the $\l$ field in higher dimensions can be interpreted as an $n-2$ form with only components in the $n$-dimensional \nc subspace of space-time. We furthermore discussed various other possibilities of implementing the Slavnov constraint(s) and also gave one version which (upon choosing an appropriate gauge fixing) features the existence of a vector supersymmetry. However, in the general $D$-dimensional case this is not sufficient to show IR finiteness of the model.

\section*{Acknowledgments}
The authors would like to thank F.~Gieres, O.~Piguet and M.~Schweda for invaluable discussions and feedback.\\
D.~N. Blaschke would like to thank the Au\ss eninstitut of Vienna University of Technology for financing a sojourn to the Theory Division at CERN in November/December 2006 during which parts of this paper were completed, as well as the Theory Division at CERN for their kind hospitality. D.~N. Blaschke is a recipient of a DOC-fellowship of the Austrian Academy of Sciences at the Institute for Theoretical Physics at Vienna University of Technology.\\
The work of Stefan Hohenegger was supported by the Austrian Bundesministerium f\"ur Wissenschaft und Forschung.

\appendix
\section{Equations of motion}\label{app:eom}
The equations of motion associated to the action (\ref{action}) are given by:
\begin{subequations}
\begin{align}
\var{\Act}{c}&=-n^iD_i\bc\,,\quad\qquad \var{\Act}{\bc}=-n^iD_ic\,,\\
\var{\Act}{\phi}&=-m^iD_i\bphi\,,\quad\qquad \var{\Act}{\bphi}=-m^iD_i\phi\,,\\
\var{\Act}{B}&=n^iA_i\,,\quad\qquad \var{\Act}{d'}=m^i\l_i\,,\\
\var{\Act}{A_i}&=D_\mu F^{\mu i}+\e^{ijk}D_j\l_k+n^i\left(B-\rig\co{\bc}{c}\right)-\rig m^i\co{\bphi}{\phi}\,,\\
\var{\Act}{A_0}&=D_kF^{k0}\,,\quad\qquad \var{\Act}{\l_i}=\inv{2}\e^{ijk}F_{jk}+m^id'\,.
\end{align}
\end{subequations}
Note, that the symmetries discussed in Section~\ref{subsec:symm} only exist if $m^i=n^i$.

\section{Propagators}\label{app-propagators}
The equations of motion associated to the bilinear part of the action (\ref{action}) including sources (and for now neglecting the ghosts) read:
\begin{subequations}\label{eom}
\begin{align}
\var{\Act_{\text{bi}}}{A^\mu}&=\square A_\mu-\partial_\mu(\partial A)+\d_\mu^i\e_{ijk}\partial^j\l^k+n_\mu B=-j_\mu^A,\label{eom1}\\
\var{\Act_{\text{bi}}}{\l^i}&=\e_{ijk}\partial^jA^k+m_id'=-j_i^\l,\label{eom2}\\
\var{\Act_{\text{bi}}}{B}&=(nA)=-j_B,\label{eom3}\\
\var{\Act_{\text{bi}}}{d'}&=(m\l)=-j_{d'}.\label{eom4}
\end{align}
\end{subequations}
By letting $\partial^\mu$ (and $\partial^i$) act on relations (\ref{eom1}) and (\ref{eom2}), respectively, one gets
\begin{align}
B&=-\frac{(\partial j_A)}{(n\partial)},\label{sol-B}\\
d'&=-\frac{(\partial j_\l)}{(m\partial)}.\label{sol-d}
\end{align}
Application of $\e_{ilm}\partial^m$ to (\ref{eom1}) then yields\footnote{In this context $\Delta\equiv\partial^i\partial_i=\square-\partial^0\partial_0$.}
\begin{align}
-\square j^\l_l+\square\frac{(\partial j_\l)}{(m\partial)}m_l+\partial_l(\partial\l)-\Delta\l_l-\e_{lmi}\partial^mn^i\frac{(\partial j_A)}{(n\partial)}=-\e_{lmi}\partial^mj_A^i,\label{rel-partl-l}
\end{align}
where equations (\ref{eom2}), (\ref{sol-B}) and (\ref{sol-d}) were inserted. Multiplying this expression with $m^l$ and using (\ref{eom4}) provides an expression for $(\partial\l)$ and after reinserting the latter into (\ref{rel-partl-l}) one finds
\begin{align}\label{sol-l}
\l_l&=\frac{\square}{\Delta}\left(-j_l^\l+\frac{(\partial j_\l)}{(m\partial)}m_l\right)+\inv{\Delta}\e_{lki}\partial^k\left(j_A^i-n^i\frac{(\partial j_A)}{(n\partial)}\right)+\nonumber\\
&\quad+\frac{\partial_l}{(m\partial)}\left[\frac{\square}{\Delta}\left((mj_\l)-m^2\frac{(\partial j_\l)}{(m\partial)}\right)-j_{d'}+\inv{\Delta}\e_{ijk}m^i\partial^j\left(\frac{(\partial j_A)}{(n\partial)}n^k-j_A^k\right)\right].
\end{align}
Finally, multiplication of (\ref{eom1}) with $n^i$ and use of equations (\ref{eom3}), (\ref{sol-B}) and (\ref{sol-l}) provides an expression for $(\partial A)$ and after reinserting the latter into (\ref{eom1}) one finds
\begin{align}\label{sol-A}
A_i=\inv{\square}\Bigg\{&-j_i^A+\frac{\partial_i}{(n\partial)}\left(\e_{jkl}n^j\partial^k\l^l-\square j_B-n^2\frac{(\partial j_A)}{(n\partial)}+(nj_A)\right)+\frac{(\partial j_A)}{(n\partial)}n_i-\nonumber\\
&-\e_{ijl}\partial^j\l^l\Bigg\},
\end{align}
where $\l^l$ is given by (\ref{sol-l}). The expression for $A_0$ is similar to (\ref{sol-A}), except for the fact that the last two terms are missing.

By varying equations (\ref{sol-B}), (\ref{sol-d}), (\ref{sol-l}) and (\ref{sol-A}) with respect to the sources and passing over to momentum space one obtains the propagators given in equations (\ref{propagators}).


%
\end{document}